# Shear Band Broadening in Simulated Glasses

*Darius D. Alix-Williams and Michael L. Falk*

LAST UPDATED - 7/5/18 2:36:08 PM

## Abstract

Two models are proposed to predict the evolution of shear band width as a function of applied strain for simulated glasses mechanically deformed in simple shear. The first model arises from dimensional analysis and an assumption that band broadening is controlled by the strain rate inside the shear band. The second model describes the shear band as a pulled front propagating into an unsteady state, the dynamics of which are described using the effective temperature shear transformation zone (ET-STZ) theory. Both models are compared to three simulated systems: a two-dimensional binary Lennard-Jones glass, a $Cu_{64}Zr_{36}$ glass modeled using an embedded atom method (EAM) potential, and a Si glass modeled using the Stillinger-Weber potential. Shear bands form in all systems across a variety of quench rates. Depending on the case these bands either appear to broaden indefinitely or to saturate to a finite width. The shear band strain rate model appears to apply only when band growth is unconstrained, indicating the dominance of a single time scale in the early stages of band development. The front propagation model, which reduces to the other model in the early stages of band broadening, also applies to cases in which the band width saturates, suggesting that competition between the rate of shear-induced configurational disordering and thermal relaxation sets a maximum width for shear bands in a variety of materials systems.

# Introduction

Strain localization, also known as shear banding, is a deformation mechanism present in a variety of glassy systems including gels [1], granular media [2] and metallic glasses [3]. Despite its ubiquitous nature, the underlying mechanisms for strain localization in these systems remain largely unknown [3]. Shear banding is often the precursor to brittle material failure in systems such as metallic glasses, which limits the utilization of these materials as structural components [3]. In spite of their brittle nature, metallic glasses often have many desirable properties such as high strength-to-weight ratios and corrosion tolerance [4]. By understanding how shear bands form, we could potentially engineer metallic glasses that avoid strain localization. Unfortunately, attempts to characterize the nature of shear bands in experimental systems have been limited by the small length and short time scales over which shear bands form [3, 5]. Simulation of shear bands presents an alternative method of studying this phenomenon.

In this study we propose two models for the width of a shear band as a function of the applied strain in simulated glassy systems driven at constant strain rate in simple shear. The shear band strain rate (SBSR) model assumes that the strain rate within the shear band is the only relevant time scale in our system and predicts that the shear band width will be proportional to the square root of strain. This model is empirical in nature, arising from dimensional analysis, and assumes a single material length scale. A second model assumes that shear bands are pulled fronts which propagate into an unsteady state. An effective temperature is used to characterize the structural state of the material ahead of and behind the front and effective temperature shear transformation zone (ET-STZ) theory defines the time-evolution of these states. We call this model the shear transformation zone pulled front (STZ-PF) model, and, in addition to assuming an existing length scale analogous to that from the SBSR model, another length scale arises in the STZ-PF model that corresponds to a shear band saturation width.

We test the applicability of the SBSR and STZ-PF models to three simulated systems – a two-dimensional binary Lennard Jones glass, a three-dimensional embedded atom method (EAM) $Cu_{64}Zr_{36}$ glass and a three-dimensional Stillinger-Weber Si glass. This paper is arranged as follows. In the Theoretical Background section, we introduce the concepts of effective temperature and shear transformation zones. We then detail the derivation of the linear spreading speed of a moving front propagating into an unsteady state based on an explicit form of the structural evolution state taken from STZ theory. In the Methods section we outline how model systems were prepared and deformed as well as the procedure used to determine the width of the shear band. In the Results section we fit our models to simulation data and comment on the fit parameters they generate. Lastly, in the Discussion section we summarize our work and comment on future directions.

## Theoretical Background

### The Shear Band Strain Rate (SBSR) Model Derivation

In deformed systems where strain localization dominates the response, a simple geometrical relationship is observed between the global strain rate, $\dot{\gamma}$, and the strain rate within the shear band, $\dot{\gamma}_b$,

$$\dot{\gamma}_b = \frac{L}{w}\dot{\gamma}.$$

(1)

In this expression $L$ is the length of the simulation cell perpendicular to the direction of the applied load and $w$ is the width of the shear band.

If one assumes that the strain rate is the dominant time scale in such simulations we can justify the existence of a simple relationship between the rate of broadening of the shear band, $\dot{w}$, and the global strain rate,

$$\dot{w} = \frac{\mathcal{L}}{2}\dot{\gamma},$$

(2)

where we introduce another length scale $\mathcal{L}$ which dictates the magnitude of broadening for a given configuration driven at constant strain rate. We surmise that this is a material length scale, with some dependence on the internal structure of the glass.

Taken together, these equations result in a simple analytic expression for the predicted width of a shear band as a function of applied strain,

$$w = \sqrt{w_0^2 + L\mathcal{L}(\gamma - \gamma_0)},$$

(3)

where $w_0$ and $\gamma_0$ are the respective band width and strain at shear band nucleation. We will refer to this model as the shear band strain rate (SBSR) model. The SBSR model suggests that, at sufficiently large applied strain, the width of the shear band is proportional to the square root of the applied strain, $w \propto \gamma^{\frac{1}{2}}$.

In the SBSR model the physics of the evolving glass is expressed using a single empirically determined parameter $\mathcal{L}$. This limits our ability to intuit response for a variety of glassy systems (metallic, covalent, etc.) or to draw connections between the response and the internal structure of a particular glass. To overcome these limitations, we seek an alternative expression for shear band broadening which incorporates the physics of ET-STZ theory, where a strong connection is drawn between material structure and response.

### Deriving the Shear-Transformation-Zone Pulled Front (STZ-PF) Model

We propose an alternative framework where band broadening is modeled as a pair of moving interfaces which separate jammed and flowing material inside and outside of the shear band. We assume that the leading edge of each interface, or front, moves outward at the linear spreading velocity $v^*$. This velocity can be determined by linearizing the dynamical equations describing the state of material ahead of the front about the unsteady state. Our approach was adapted from a review of front propagation dynamics by W. van Saarloos [6].

In his review, van Saarloos defines *pulled fronts* as the class of interfaces whose propagation speed is exactly $v^*$ in steady-state. The speed of a growing and spreading perturbation, $u(x,t)$, can be characterized by constructing a level set line through this curve at some fixed, arbitrary value $u(x,t) = C$. The time-rate-of-change of the position of the intersection point of the line and the curve, $\frac{dx_c(t)}{dt}$, is expected to reach an asymptotic velocity, $v^*$ as $t \to \infty$. This velocity can be calculated by linearizing the dynamical equations describing the structural state of the material ahead of the front and using a saddle-point approximation.

The details of this procedure are discussed extensively in Ref [6]. In this subsection we first introduce effective temperature and shear transformation zones. These concepts are then combined in a dynamical model that describes the evolution of the structural state of our glassy systems. We linearize this model about a spatiotemporal perturbation to obtain a dispersion relation. The dispersion relation is analyzed at the saddle-point to find the linear spreading velocity, which is taken to determine the rate of band broadening. Finally, we find an analytic expression for the shear band width as a function of strain using this rate.

The concept of an effective temperature stems from the idea that non-equilibrium systems can be thought of as consisting of two, weakly-coupled subsystems [7]. The first subsystem is what we traditionally imagine when referring to temperature and encompasses the fast kinetic/vibrational degrees of freedom that quickly come into equilibrium with the surroundings, while the second subsystem refers to the slow configurational degrees of freedom in the system. If one imagines a vitreous system surrounded by a reservoir that is quickly cooled from a high temperature above its melting point down to a temperature far below its glass transition temperature, at some point the slow sub-system and reservoir will fall out of equilibrium with each other, leaving the slow sub-system in a configuration on the potential energy landscape drawn from an ensemble typical of the glass transition temperature [8], often referred to as the fictive temperature. If we repeat this procedure at a slower cooling rate the system and reservoir remain in thermal equilibrium with each other for longer and fall out of equilibrium at a lower temperature. The slow sub-system of this second glass will have a configuration drawn from this lower temperature ensemble and will thus have a higher degree of structural order and a lower effective temperature.

In the effective temperature shear transformation zone (ET-STZ) theory it is proposed that when one does plastic work on such a glassy system nanoscopic defects called shear transformation zones (STZs) undergo shear-induced structural rearrangements dissipating energy. A typical STZ contains 10s of atoms [3]. A simple model of STZs assumes that they have two states and, once rearranged, cease further transformation unless the direction of shear is reversed. For systems driven at constant strain rate, new STZs must be created via the dissipated plastic work to sustain the flow.

ET-STZ theory has undergone numerous changes and appeared in various forms since its introduction [8, 9, 10]. Our simulations are performed at finite temperature therefore we desire a dynamical equation for the dimensionless effective temperature, $\chi$, which includes the effects of shear-induced rejuvenation and thermal relaxation. We consider the shear banded system as spatially invariant in all but the y-dimension, reducing shear band broadening to a spatially one-dimensional problem. At lower strain shear bands can be spatially heterogeneous as evinced by various topological features and potential energy fluctuations. Shear bands become increasing homogeneous at higher strain, suggesting that effective temperature transport occurs in driven glassy systems. The dynamical equation for the evolution of the structural state should include a diffusive term to account for this effect.

The general structure for the evolution of the effective temperature proposed by Manning *et al.* in Ref [11] Equation A5 is

$$\dot{\chi} = \frac{1}{C_{eff}T_Z}\left\{T_{eff}\left(\frac{dS_C}{dt}\right)_{mech}\left[1 - \frac{\chi}{\hat{\chi}}\right] + T_{eff}\left(\frac{dS_C}{dt}\right)_{therm}\left[1 - \frac{\chi T_Z}{T}\right]\right\} + D\frac{\partial^2 \chi}{\partial y^2}.$$

(4)

In this expression $C_{eff}$ is a specific heat and $T_Z = E_Z/k_B$, where $E_Z$ is the energy required to nucleate a STZ and $k_B$ is the Boltzmann constant.

Heat due to mechanical work done on the system, $T_{eff}(dS_c/dt)_{mech}$, drives the structural state, $\chi = T_{eff}/T_Z$, towards the steady-state value $\hat{\chi}$. The existence of $\hat{\chi}$ has been demonstrated in a prior study of the LJ system [12] where a linear relation was assumed between effective temperature and potential energy and the average potential energy of material within the shear band was found to converge to a fixed value. Conversely, heat generated from thermal fluctuations, $T_{eff}(dS_c/dt)_{therm}$, relaxes the structural state towards the bath temperature T. We introduce a dimensionless bath temperature $\theta = T/T_Z$, analogous to the dimensionless effective temperature. A final term allows structural disorder to diffuse. We assume the rate of diffusion is governed by the plastic strain rate $\dot{\gamma}$ yielding a coefficient, $D = \ell^2|\dot{\gamma}|$, where $\ell$ is a length scale on the order of a STZ radius. The equation for the plastic strain rate is

$$\dot{\gamma} = \frac{2}{\tau_0} f(s) e^{-1/\chi}.$$

(5)

The parameter $\tau_0$ is an internal time scale comparable to the phonon frequency and $f(s)$ is a function of the deviatoric shear stress, $s$. The plastic strain rate is the product of the rate of stress induced STZ transitions, $2f(s)/\tau_0$, and an Arrhenius term that is proportional to the number density of STZs, $\exp(-1/\chi)$. It is assumed that STZ transitions occur only in the direction of loading. We hold off providing an explicit definition of $f(s)$ until later in our derivation.

In Ref [11], Eq. A13 Manning *et al.* provide an expression for the rate of configurational entropy production due to mechanical loading,

$$\left(\frac{dS_C}{dt}\right)_{mech} = \frac{k_B v_Z}{\Omega} \frac{\epsilon_0}{\tau_0} \Lambda \Gamma(s),$$

(6)

where $v_Z$ is the number of molecules within an STZ, $\Omega$ is the volume per molecule, $\epsilon_0$ is a strain increment of order unity, $\Lambda = \exp(-1/\chi)$ is proportional to the STZ density and $\Gamma(s)$ is the energy dissipated per STZ.

The STZ energy dissipation term $\Gamma(s)$ (Eq. A10) is given as

$$\Gamma(s) = \frac{2}{s_0 \epsilon_0} s f(s),$$

(7)

where $s_0$ is the minimum flow stress and $s$ is the deviatoric shear stress.

Combining Equations (6) and (7),

$$\left(\frac{dS_C}{dt}\right)_{mech} = \frac{k_B v_z}{\Omega} \frac{\epsilon_0}{\tau_0} e^{-1/\chi} \frac{2}{s_0 \epsilon_0} s f(s).$$

(8)

Manning *et al.* analyze strain localization simulations performed at low temperature and assume that the contribution due to thermal fluctuations is marginal. Consequently, an expression for $(dS_c/dt)_{therm}$ is not provided. We refer instead to Ref [10] Eq. 6.2 where Langer suggests the following form

$$\left(\frac{dS_C}{dt}\right)_{therm} = \kappa \frac{k_B v_z}{\Omega} \frac{\epsilon_0}{\tau_0} \rho(T) e^{-\beta/\chi}.$$

(9)

In this expression $\kappa$ is a dimensionless scaling parameter, $\rho(T)$ is a thermal factor whose form is beyond the scope of this derivation, and $\beta$ is an activation term which dictates the susceptibility of STZ transitions to thermal fluctuations.

Manning *et al.* define a dimensionless effective temperature $\tilde{c}_0 = C_{eff} \Omega/(k_B v_z)$. Combining Equations (4) - (9), the evolution of the structural state is expressed as

$$\dot{\chi} = \frac{\chi \epsilon_0}{\tilde{c}_0 \tau_0} \left\{ \frac{2}{s_0 \epsilon_0} s f(s) e^{-1/\chi} \left[1 - \frac{\chi}{\hat{\chi}}\right] + \kappa \rho(T) e^{-\beta/\chi} \left[1 - \frac{\chi}{\theta}\right] \right\}$$
$$+ \ell^2 \frac{2}{\tau_0} f(s) e^{-1/\chi} \frac{\partial^2 \chi}{\partial y^2}.$$

(10)

We know from prior studies [11] that this equation is unstable to strain localization due to a non-linear instability. However, the question we ask now is, "Once this instability takes place, how does the resulting band broaden as material adjacent to the band is induced to flow along with the material in the initially nucleated band?" To answer this question, we refer to the literature on the propagation of fronts into unstable states [6], noting that the jammed material outside the band is unstable in the ET-STZ model. If we assume the propagation is consistent with a pulled front, the unstable spreading speed can be obtained by analyzing the linear stability of the unstable state, i.e. the material outside the band. To undertake this analysis, we decompose the structural state of the system far from the band into a spatially invariant term, $\chi_0(t)$, and some small perturbation $u(y, t)$ where $u \ll \chi_0$,

$$\chi(y, t) \equiv \chi_0(t) + u(y, t).$$

(11)

We give the perturbation the following form

$$u = u_0 \exp(iky - i\omega t).$$

(12)

This is a generic plane wave with amplitude $u_0$, wavenumber $k$, and angular frequency $\omega$. Parameters $y$ and $t$ represent position and time, respectively. The plane wave equation is in complex exponential form with imaginary number $i$. The expression for the evolution of the uniform part of the solution is

$$\dot{\chi}_0 = \frac{\chi_0 \epsilon_0}{\tilde{c}_0 \tau_0} \left\{ \frac{2}{s_0 \epsilon_0} sf(s) e^{-1/\chi_0} \left[1 - \frac{\chi_0}{\hat{\chi}}\right] + \kappa \rho(T) e^{-\beta/\chi_0} \left[1 - \frac{\chi_0}{\theta}\right] \right\}.$$

(13)

Since $u \ll \chi_0$, we can make the approximations,

$$\exp\left(\frac{-1}{\chi_0 + u}\right) \approx \exp(-1/\chi_0)\left[1 + \frac{u}{\chi_0^2}\right],$$

(14)

$$\exp\left(\frac{-\beta}{\chi_0 + u}\right) \approx \exp(-\beta/\chi_0)\left[1 + \frac{\beta u}{\chi_0^2}\right].$$

(15)

We now solve for the linear solution of $\dot{u}$, excluding any terms of order $O(\chi_0^{-2})$ or higher.

$$\dot{u} = \frac{\epsilon_0}{\tilde{c}_0 \tau_0}\left\{\frac{2}{s_0 \epsilon_0} sf(s)e^{-1/\chi_0}\left(1 + \frac{1}{\chi_0} - \frac{1}{\hat{\chi}} - 2\frac{\chi_0}{\hat{\chi}}\right) + \kappa\rho(T)e^{-\beta/\chi_0}\left(1 + \frac{\beta}{\chi_0} - \frac{\beta}{\theta} - 2\frac{\chi_0}{\theta}\right)\right\} u$$
$$+ \ell^2 \frac{2}{\tau_0} f(s) e^{-1/\chi_0} \frac{\partial^2 u}{\partial y^2}$$

(16)

We introduce a dimensionless rate factor $\alpha$ that determines the rate at which the perturbation grows or decays due to mechanical work,

$$\alpha = \frac{2}{\tilde{c}_0}\left(1 + \frac{1}{\chi_0} - \frac{1}{\hat{\chi}} - 2\frac{\chi_0}{\hat{\chi}}\right)e^{-1/\chi_0}.$$

(17)

Recall that $\chi_0$ characterizes the structural state of jammed material. There is a critical effective temperature $\chi_c$ beneath which a perturbation is expected to grow, i.e. for $\chi_0 < \chi_c < \hat{\chi}$. When the structural state of the jammed material exceeds this critical value, $\chi_0 > \chi_c$, a perturbation will decay. $\chi_c$ represents the crossover from heterogeneous to homogeneous deformation when thermal relaxation is ignored.

We introduce a second dimensionless rate factor that determines the direction of perturbation growth due to thermal relaxation,

$$\tilde{\kappa} = -\kappa\rho(T)\frac{\epsilon_0}{\tilde{c}_0}\left(1 + \frac{\beta}{\chi_0} - \frac{\beta}{\theta} - 2\frac{\chi_0}{\theta}\right)e^{-\beta/\chi_0}.$$

(18)

The term in parenthesis is negative because $\theta \leq \chi_0 \leq \beta$, while all other terms are positive. We introduce a minus sign to yield a positive rate factor. A sign change also occurs in the perturbation growth rate equation $\dot{u}$ and indicates that thermal relaxation damps perturbations. We further assume that the stress in our systems is equal to the minimum flow stress, $s = s_0$. Substitution of Equations (17) and (18) along with our assumed stress state into (16) results in the following expression for the time evolution of the perturbation

$$\dot{u} = \frac{1}{\tau_0}(\alpha f(s_0) - \tilde{\kappa})u + \ell^2 \frac{2}{\tau_0} f(s_0) e^{-1/\chi_0} \frac{\partial^2 u}{\partial y^2}.$$

(19)

Recall that $u = u_0 \exp(iky - i\omega t)$. We can obtain the dispersion relation

$$\omega = \frac{i}{\tau_0}\left[\alpha f(s_0) - \tilde{\kappa} - \ell^2 \frac{2}{\tau_0} f(s_0) e^{-1/\chi_0} k^2\right],$$

(20)

where the spatial wavenumber $k$ is complex, and its real and imaginary parts are denoted using the subscripts $r$ and $i$, respectively.

As discussed at length in Ref. [6], the linear spreading speed of the interface can be extracted from the dispersion relation via a saddle-point approximation. The k-value of the saddle-point, $k^*$, and the linear spreading speed, $v_0^*$, are then given by Eq. 12 of Ref. [6],

$$\left.\frac{d\omega}{dk}\right|_{k^*} = \frac{\omega_i(k^*)}{k_i^*},$$

(21)

$$v_0^* = \frac{\omega_i(k^*)}{k_i^*}.$$

(22)

Equation 21 can be separated into its real and imaginary components and used to determine the critical wave number,

$$k^* = i\sqrt{\frac{\alpha f(s_0) - \tilde{\kappa}}{\ell^2\, 2f(s_0)e^{-1/\chi_0}}}.$$

(23)

The linear spreading speed is therefore

$$v_0^* = 2\sqrt{2}\ell\frac{f(s_0)}{\tau_0}\sqrt{\alpha e^{-1/\chi_0}}\sqrt{1 - \frac{\tilde{\kappa}}{\alpha f(s_0)}}.$$

(24)

We assume that the STZ activation rate is proportional to the strain rate within the shear band,

$$f(s_0) \sim \frac{\tau_0}{2}\frac{V}{w}e^{1/\hat{\chi}},$$

(25)

where $V$ is the constant velocity imposed on the simulation cell at the boundary, $w$ is the width of the shear band, and the density of STZs within the shear band is assumed constant and given by $exp(-1/\hat{\chi})$. The rate of band broadening is therefore

$$\dot{w} = 2v_0^* = 2\sqrt{2}\ell e^{1/\hat{\chi}}\sqrt{\alpha e^{-1/\chi_0}}\frac{V}{w}\sqrt{1 - \frac{2\tilde{\kappa}w}{\alpha\tau_0 e^{1/\hat{\chi}}V}}.$$

(26)

We define the dynamic length scale

$$\mathcal{L} = 4\sqrt{2}\ell e^{1/\hat{\chi}}\sqrt{\alpha e^{-1/\chi_0}},$$

(27)

that governs the initial rate of band broadening at low strain when $w \ll w_\infty$. This length scale appears in the SBSR model and is now explicitly defined in terms of the ET-STZ theory. To understand the behavior of this length scale, we first expand this expression by substitution of Equation (17) to obtain

$$\mathcal{L} = 8\ell e^{1/\hat{\chi}-1/\chi_0}\sqrt{\frac{1 + \frac{1}{\chi_0} - \frac{1}{\hat{\chi}} - 2\frac{\chi_0}{\hat{\chi}}}{\tilde{c}_0}}.$$

(28)

The exponential term is the proportion of STZs within the jammed material at the maximum allowed disorder due to amorphous packing and approaches 1 as $\chi_0 \to \hat{\chi}$. As the jammed material becomes more disordered, the radical term decreases, suggesting a reduction in the energy dissipated into the configurational degrees of freedom. The combined effect is a dynamic length scale which increases the shear band growth rate when the state of jammed material is more disordered up to some critical value $\chi_0 \to \chi_C$. At even higher levels of disorder the shear band growth rate sharply decreases until reaching zero at the degree of disordering above which shear bands never form. The functional form of $\mathcal{L}$ thus suggests that shear bands typically broaden faster in systems with more disordered states.

We can also define a band width saturation length scale,

$$w_\infty = \frac{1}{2}\frac{\alpha \tau_0 V}{\tilde{\kappa}} e^{1/\hat{\chi}},$$

(29)

which determines when the rate of shear band broadening goes to zero. The dependence on both rate factors $\alpha$ and $\tilde{\kappa}$ suggests that this length scale is set by the competition between stress induced structural rejuvenation and thermal relaxation. We explore the behavior of the saturation length scale by expanding Eqn. (29),

$$w_\infty = \frac{\tau_0 V}{\epsilon_0 \kappa \rho(T)} \left( \frac{1 + \frac{1}{\chi_0} - \frac{1}{\hat{\chi}} - 2\frac{\chi_0}{\hat{\chi}}}{2\frac{\chi_0}{\theta} + \frac{\beta}{\theta} - \frac{\beta}{\chi_0} - 1} \right) \exp\left(\frac{1}{\hat{\chi}} - \left[\frac{1-\beta}{\chi_0}\right]\right).$$

(30)

The term in the parenthesis decreases as the effective temperature of the jammed state, $\chi_0$, increases, once again demonstrating that thermal relaxation dampens the effects of shear induced disordering resulting in the reduction of the saturation length scale. The exponential term, by contrast, represents the fraction of STZs available to deform, which increases with effective temperature. The combined effect is a saturation length scale that is small for less disordered systems with low effective temperatures but which increases exponentially for systems whose disorder temperature approaches the steady state value.

By substitution of Eqns. (27) and (29) into (26), we arrive at the following expression for the band broadening rate

$$\dot{w} = \frac{\mathcal{L}V}{2w}\sqrt{1 - \frac{w}{w_\infty}}.$$

(31)

From this expression the aforementioned effects of $\mathcal{L}$ and $w_\infty$ are apparent. When $w \ll w_\infty$, the rate of band broadening is proportional to $\mathcal{L}$ and inversely proportional to $w$ and narrow bands are expected to grow faster than wider ones. As the band width approaches its saturation value, $w \to w_\infty$, the band growth rate is reduced to zero as, $\dot{w} \to 0$.

The analytic solution to Eq. (31) is

$$\left(2 + \frac{w}{w_\infty}\right)\sqrt{1 - \frac{w}{w_\infty}} - \left(2 + \frac{w_0}{w_\infty}\right)\sqrt{1 - \frac{w_0}{w_\infty}} = \frac{3}{4}\frac{\mathcal{L}V}{w_\infty^2}(t - t_0) = \frac{3}{4}\frac{\mathcal{L}L}{w_\infty^2}(\gamma - \gamma_0).$$

(32)

In this expression, $w_0$ and $\gamma_0$ are the band width and strain at shear band nucleation. We refer to this expression as the shear transformation zone pulled front (STZ-PF) model. In the limit where $w_0 \ll w_\infty$ and $w \ll w_\infty$,

$$w^2 \approx w_0^2 + \mathcal{L}L(\gamma - \gamma_0),$$

(33)

the same functional form as the SBSR model.

## Methods

### System Preparation

The shear band broadening models are tested on simulated glasses generated using LAMMPS [14]. Three systems are chosen to represent metallic and covalent glasses with planar geometry and correspond to a two-dimensional binary Lennard-Jones glass (LJ) [8], a three-dimensional $Cu_{64}Zr_{36}$ glass (CZ) modeled using an Embedded Atom Method potential [13] and a three-dimensional silicon glass (Si) modeled with a Stillinger-Weber potential [15]. We restrict the depth of the CZ and Si systems in the z-dimension such that they are effectively two-dimensional. This simulation box depth is chosen to be larger than the cutoff of the respective interatomic potentials, as evinced by plotting the radial distribution functions (not shown) and observing that pair interactions are uncorrelated at this length. All simulations have a total of 80,000 atoms, an aspect ratio $L_x:L_y$ of 1:5 and periodic boundaries in all directions. Shear bands form in all systems when deformed in simple shear.

The LJ system was introduced by Lançon *et al.* in a study of phase stability of simulated quasi-crystals [16]. We modify this system to include a pairwise interaction cutoff distance of 2.5 $\sigma$. The length of the simulation cell perpendicular to the loading direction, $L$, is 639.5 $\sigma$. There are 35,776 large and 44,224 small particles. We prepare three LJ configurations via a constant volume quench from a well-equilibrated liquid at 0.351 $k_B/\epsilon$ to 0.0299 $k_B/\epsilon$ over durations of 1,000 $\tau$ (LJ-1), 10,000 $\tau$ (LJ-2) and 100,000 $\tau$ (LJ-3). Temperature is controlled using a Nose-Hoover thermostat. We prepare ten replicas for each configuration to minimize sample-to-sample variation. The internal structures of each LJ configuration are statistically different as indicated by their respective average potential energies: -2.15 $\epsilon$ (LJ-1), -2.17 $\epsilon$ (LJ-2) and -2.19 $\epsilon$ (LJ-3).

Preparation of the CZ system follows the work of Shimizu *et al.* [17]. We use an aspect ratio $L_x : L_y : L_z$ of 4:20:1 with $L_y = L = 502.98$ Å. Our system is composed of 28,800 Zr and 51,200 Cu atoms. Three CZ configurations are prepared through constant pressure quench from 2000 to 300 K at rates of 1.00 (CZ-1), 0.10 (CZ-2) and 0.01 (CZ-3) K/ps. Temperature and pressure are controlled using a Nose-Hoover thermostat and barostat. Five replicas of each configuration are prepared, with the average atomic energy of Cu atoms of -3.569 (CZ-1), -3.576 (CZ-2) and -3.589 (CZ-3) eV.

The final system is a silicon glass whose preparation was first introduced by Fusco *et al.* [15]. This system has a 4:20:1 aspect ratio with $L = 547.81$ Å and 80,000 atoms. We prepared three configurations by quenching from 3500 K to 300 K at constant pressure using the Tersoff potential and a Nose-Hoover thermostat and barostat. The quench rates were 0.100 K/ps (Si-1), 0.010 K/ps (Si-2) and 0.001 K/ps (Si-3). The system is subsequently annealed at 400 K for 100 ps using the Stillinger-Weber potential. Five replicas of each Si configuration are prepared with average potential energy values of -4.09 (Si-1), -4.10 (Si-2) and -4.11 (Si-3) eV.

### Deformation

We deform our systems in simple shear by incrementally deforming the simulation cell at a constant rate of 0.0001 ps$^{-1}$ ($\tau^{-1}$, LJ). We integrate the SLLOD equations of motion [18] [19] with fully periodic boundaries and velocity remapping for atoms which cross the periodic boundaries in the y-direction. This shear protocol is consistent with Lees-Edwards boundary conditions. Strain localization occurs in all simulations; however, the chosen geometry minimizes stress concentration allowing us to strain the systems in excess of 1000%. The systems are coupled to a Nose-Hoover thermostat at 300 K (CZ and Si) and 0.0299 $k_B/\epsilon$ (LJ) to remove heat generated during shear.

## Measuring Shear Band Width

We compute the local atomic strain using OVITO [20] using the initial, undeformed configuration as the reference state. A cutoff radius of 6 Å (CZ and Si) or 2.2 $\sigma$ (LJ) is used to establish the neighborhood of each atom. The neighbor cutoff is chosen as roughly twice the distance to the highest peak of the radial distribution function. We interpolate the atomic strain onto a square grid of length 1 Å (CZ and Si) or 1 $\sigma$ (LJ) using a natural method [21]. We then apply a binary mask to the grid which labels any square with an average atomic strain greater than or equal to 0.25 as one and all others zero. A feature is defined as a cluster of adjacent ones and may cross the upper or lower cell bounds in the y-direction. We assume that the largest feature is the shear band and its width is measured by the average number of adjacent squares along the y-direction. The initial shear band width is measured at 20% strain then again from 100% and onward at 100% strain increments. In all simulations a shear band has formed by 20% strain. Simulations are excluded from subsequent analysis if additional system-spanning features are present at any stage of the deformation, which would indicate the formation of a secondary shear band.

## Results

The relative stability of a glass is measured by its average atomic energy and the likelihood that multiple shear bands will form increases with decreasing glass stability. Table 1 lists the average atomic energy and number of simulations considered for each configuration. For all systems, the 1$^{st}$ configuration is the least stable while the 3$^{rd}$ configuration is the most stable due to its lower relative mean atomic energy. Consider the LJ system where 10 simulations are performed for each configuration. As LJ configurations are generated at slower quench rates, the mean atomic energy decreases and fewer simulations are excluded due to the formation of secondary shear bands. This trend is also observed in the CZ and Si systems, but secondary shear bands are more infrequent due to the need for instabilities to traverse both the width and depth of the three-dimensional simulation cell.

The SBSR and STZ-PF models assume that shear bands have well-defined edges and constant width; however, actual shear bands have more complex structures. Figure 3 shows several features for representative CZ (a, b) and LJ (c, d) simulations in black. These features correspond to regions where the average atomic strain is greater than or equal to 25% when the system is deformed to 20% shear strain (top row) and 100% strain (bottom row). The shear band shown in the CZ configuration (a) represents the ideal case where a single shear band forms and remains the dominant feature throughout the study.

In (b) a secondary shear band is present at 20% strain and broadening occurs in both features. Secondary bands may form at any stage of deformation and we exclude simulations from our analysis if they do. The formation of additional features indicates that in addition to aging, plastic deformation occurs in material outside the primary shear band. The STZ-PF model incorporates these effects by including a thermal relaxation term, $\kappa$ and a structural rejuvenation term that depends on the local strain rate $\dot{\gamma}$.

Figure 3 (c, d) shows shear bands with poorly defined edges due to intermixed regions of low strain. These pore-likefeatures occur most often in the two-dimensional LJ simulations at early stages of shear band formation ($\gamma \sim 20\%$) but are less pronounced at later stages of deformation. The coexistence of low and high-strain regions within shear bands is less prevalent in CZ and Si simulations, which may be result of averaging strain over the third dimension.

Vertical shear bands may occur at early stages of deformation as shown by the tail of the dominant feature in Figure 3 (d). These features persist in our simulations at 100% strain and dictate the preferred direction of band broadening. Vertical banding suggests that a stress bias is not well established immediately after yielding, allowing initial STZ transitions to occur in all directions. As a result, shear bands can form in both the horizontal and vertical directions. As the system is driven to higher strain a stress bias begins to dominate the response, causing STZ transitions to occur primarily in the loading direction. The geometry of our simulation cell was chosen to minimize these effects as a high aspect ratio favors horizontal banding.

Two scenarios are possible for simulations where a single shear band dominates the response: (1) unconstrained growth, where the shear band eventually engulfs the entire cell and the simulation fluidizes, and (2) shear band saturation, where the shear band ceases to spread to the simulation cell extents and coexisting jammed and unjammed material regions exist upon further strain.

Figure 4 plots band width as a function of global strain for the Si-1 configuration. Fits are performed for each simlation and average model parameter values are used to plot the STZ-PF (dashed line) and SBSR (dotted line) model results. Pane (a) averages over simulations where the STZ-PF model predicts unconstrained band growth. In this limit, $w_\infty$ approaches infinity and the STZ-PF model reduces to the SBSR model and the predicted fits coincide. Model fits are compared for simulations where shear bands saturate in Pane (b). The STZ-PF model is far better at capturing the behavior of the Si-1 system at the low and high-strain limits.

We use R-squared ($R^2$) as a metric for comparing the goodness of each model. We exclude simulations where unconstrained growth is determined by the STZ-PF model because $w_\infty$ is undefined. Table 3 provides average coefficient data as predicted by each model. This data is averaged over N configurations and in all cases the STZ-PF model provides marked improvement over the SBSR model. The SBSR model tends to under-approximate the dynamic length scale $\mathcal{L}$, which effectively determines the band width at low strain.

Figure 5 compares these models on the LJ-1 configuration where all simulations show constrained band broadening. The slope of the SBSR curve at early strain is much smaller than that of the STZ-PF model. This is the result of the SBSR model having to strike a balance between accounting for the rapid increase in band width at low strain and the subsequent decrease and eventual plateau in band width at high strain. The addition of the saturation length scale in the STZ-PF model significantly enhances the estimation of band width and for the LJ-1 configuration the $R^2$ value increases from 0.84 (SBSR) to 0.96 (STZ-PF).

Shear band saturation is a ubiquitous phenomenon in the systems studied and occurs in the majority of simulations. In Figure 6 we plot shear band width normalized by the height of the simulation cell as a function of applied strain for the LJ (a), CZ (b) and Si (c) systems. Shapes correspond to configurations of a given system with circles indicating the fastest quench and triangles the slowest. Averages are taken over simulations where the STZ-PF model suggests band saturation with standard deviation bars shown. We assume that shear bands initiate at 20% strain and find that the initial width and dynamic length scale of the shear bands increases with increasing quench rate. The difference in initial band widths is less pronounced in the Si system where values are identical to within measured deviation. Saturation length also varies with quench rate with a larger width predicted for quickly-quenched systems.

## Discussion

This study compares two models of shear band broadening - the shear band strain rate (SBSR) model which assumes that the rate of band broadening is proportional to the strain rate within the shear band and the shear transformation zone pulled front (STZ-PF) model that describes shear band broadening as a pulled front propagating into an unsteady state. We test these models on three systems: a two-dimensional Lennard-Jones glass, a three-dimensional embedded atom method $Cu_{64}Zr_{36}$ glass and a three-dimensional Stillinger-Weber silicon glass. Shear bands form in each system and their dynamics are well captured by the STZ-PF model, even when configurations are generated at increasingly high cooling rates.

Initially, shear bands grow at a rate proportional to the square root of applied strain. Shear bands may grow indefinitely with increasing strain, but more often saturate to a finite length forming coexisting regions of jammed and flowing material. The SBSR model is able to capture the dynamics of shear bands at the onset of strain localization or whenever band width is unconstrained; however, the SBSR model breaks down as shear bands saturate to finite width. The STZ-PF model improves upon this shortfall in the SBSR model by introducing a band width saturation length scale that depends on the internal structure and rate of thermal relaxation of the glass. The STZ-PF model reduces to the SBSR model whenever band width is much smaller than saturation length, such as the case of unconstrained band growth.

This work represents the first step in an ongoing study of strain localization in simulated glasses. It demonstrates that the structural state of sheared glassy systems can be well-described using an effective temperature (ET) and the dynamics of the state can be described using shear transformation zone (STZ) theory. Our results suggest that shear bands can be modeled as pulled fronts which propagates into an unsteady state. The dynamics of the state on either side of the front can be described as a competition between increasing structural disorder due to plastic work and increasing structural order due to thermal relaxation.

Experimental studies of shear banding in metallic glasses report widths ranging from 10 – 210 nm [22] while our STZ-PF model predicts band widths between 6.9 (CZ-3) and 34.7 (CZ-1) nm.

Although these results seem promising, we acknowledge several difficulties when drawing comparisons between simulated and experimental systems. We deform our simulations in simple shear while experiments tend to compress samples using nanoindentation [22]. The simplified geometry of our simulations allows them to deform to strains typically unobservable in experimental systems, where stress concentration often leads to catastrophic failure before a steady-state flow stress is achieved. Our simulations are effectively two-dimensional, and our analysis is done in one dimension. In contrast, experimental systems have finite depth and bands must broaden both perpendicular to the loading direction and along this depth. The length and time scales of simulations and experiments vary by orders of magnitude. Experimental systems have length scales of millimeters and time scales of milliseconds while our simulations are on scales of nanometers and femtoseconds, respectively.

This preliminary study leaves many questions unanswered. A subsequent investigation will focus on understanding the role of thermally-induced structural relaxation by shearing LJ glasses at different bath temperatures. We predict that as the reservoir temperature increases, thermal relaxation should play a larger role in the material response leading to a decrease in saturation length. This study should provide insight into the functional form of $\kappa$. Several model parameters depend explicitly on the structural state of the system, which is poorly defined. We also hope to elucidate the relationship between potential energy and effective temperature in order to quantify $\chi$ and completely define our systems in terms of ET-STZ theory.

This work was supported through NSF award DMR 1408685/1409560. The authors would also like to acknowledge helpful conversations with James Langer, Robert Maaß, Sylvain Patinet and Mark Robbins.

# References

[1] M. Oda, T. Takemura and M. Takahashi, "Microstructure in shear band observed by microfocus X-ray computed tomography," *Geotechnique,* vol. 54, no. 8, pp. 539-542, 2004.

[2] A. R. Hinkle and M. L. Falk, "A small-gap effective-temperature model of transient shear band formation during flow," *Journal of Rheology,* vol. 60, no. 873, 2016.

[3] R. Maaß and J. F. Löffler, "Shear-Band Dynamics in Metallic Glasses," *Advanced Functional Materials,* vol. 25, pp. 2353-2368, 2015.

[4] M. M. Trexler and N. N. Thadhani, "Mechanical properties of bulk metallic glasses," *Progress in Materials Science,* vol. 55, pp. 759-839, 2010.

[5] T. C. Hufnagel, C. A. Schuh and M. L. Falk, "Deformation of metallic glasses: Recent developments in theory, simulations, and experiments," *Acta Materialia,* pp. 375-393, 2016.

[6] W. van Saarloos, "Front propagation into unstable states," *Physics Reports,* vol. 386, pp. 29-222, 2003.

[7] M. L. Falk and J. S. Langer, "Deformation and Failure of Amorphous, Solidlike Materials," *Annual Review of Condensed Matter Physics,* vol. 2, pp. 353-373, 2011.

[8] Y. Fan, T. Iwashita and T. Egami, "Energy landscape-driven non-equilibrium evolution of inherent structure in disordered material," *Nature Communications,* vol. 8, no. 15417, 2017.

[9] M. L. Falk and J. S. Langer, "Dynamics of viscoplastic deformation in amorphous solids," *Physical Review E,* vol. 57, no. 6, 1998.

[10] J. S. Langer, "Dynamics of shear-transformation zones in amorphous plasticity: Formulation in terms of an effective disorder temperature," *Physical Review E ,* vol. 70, no. 041502, 2004.

[11] J. S. Langer, "Shear-transformation-zone theory of plastic deformation near glass transition," *Physical Review E,* vol. 77, no. 021502, 2008.

[12] M. L. Manning, E. G. Daub, J. S. Langer and J. M. Carlson, "Rate-dependent shear bands in a shear-transformation-zone model of amorphous solids," *Physical Review E,* vol. 79, no. 016110, 2009.

[13] Y. Shi, M. B. Katz, H. Li and M. L. Falk, "Evaluation of the Disorder Temperature and Free-Volume Formalisms via Simulations of Shear Banding in Amorphous Solids," *Physical Review Letters,* vol. 98, no. 185505, 2007.

[14] M. S. Daw and M. I. Baskes, "Embedded-atom method: Derivation and application to impurities, surfaces, and other defects in metals," *Physical Review B,* vol. 29, no. 6443, 1984.

[15] S. Plimpton, "Fast Parallel Algorithms for Short-Range Molecular Dynamics," *J Comp Phys,* vol. 117, pp. 1-19, 1995.

[16] C. Fusco, T. Albaret and A. Tanguy, "Role of local order in the small-scale plasticity of model amorphous materials," *Phys. Rev. E,* vol. 82, no. 066116, 2010.

[17] F. Lançon, L. Billard and P. Chaudhari, "Thermodynamical Properties of a Two-Dimensional Quasi-Crystal from Molecular Dynamics Calculations," *Eurphysics Letters,* vol. 2, pp. 625-629, 2007.

[18] F. Shimizu, S. Ogata and J. Li, "Theory of Shear Banding in Metallic Glasses and Molecular Dynamics Calculations," *Materials Transactions,* vol. 48, no. 11, pp. 2923-2927, 2007.

[19] D. J. Evans and G. P. Morriss, "Nonlinear-response theory for steady planar Couette flow," *Physical Review A,* vol. 30, no. 1528, 1984.

[20] P. J. Daivis and B. D. Todd, "A simple, direct derivation and proof of the validity of the SLLOD equations of motion for generalized homogeneous flows," *The Journal of Chemical Physics,* vol. 124, no. 194103, 2006.

[21] A. Stukowski, "Visualization and analysis of atomistic simulation data with OVITO - the Open Visualization Tool," *Modeling Simul. Mater. Sci. Eng.,* vol. 18, no. 015012, 2010.

[22] I. Amidror, "Scattered data interpolation methods for electronic imaging systems: a survey," *Journal of Electronic Imaging,* vol. 11, no. 2, p. 157, 2002.

[23] C. Liu, V. Roddatis, P. Kenesei and R. Maaß, "Shear-band thickness and shear-band cavities in a Zr-based metallic glass," *Acta Materialia,* vol. 140, pp. 206-216, 2017.

[24] D. E. a. G. Morris, Statistical Mechanics of Nonequilibrium Liquids, New York, 1990.

# Tables

Table 1 – Simulation details for representative LJ configurations. A total of 10 simulations were conducted for each configuration; however, simulations were excluded from the study if secondary shear bands formed during deformation.

| Configuration Name | Quench Duration ($\tau$) | Average Atomic Energy ($\epsilon$) | Number of Simulations |
|---|---|---|---|
| Lennard Jones Potential | | | |
| **LJ-1** | 1e4 | -2.15 ± .13 | 6 |
| **LJ-2** | 1e5 | -2.17 ± .12 | 7 |
| **LJ-3** | 1e6 | -2.19 ± .12 | 10 |

Table 2 – Simulation details for representative CZ and Si configurations. Five (5) simulations were conducted for each configuration; however, simulations are excluded from the study if secondary shear bands form during deformation.

| Configuration Name | Quench Rate (K·ps) | Average Atomic Energy (eV) | Number of Simulations |
|---|---|---|---|
| Sheng Embedded Atom Method Potential | | | |
| **CZ-1** | 0.100 | $-3.57 \pm .07$ | 3 |
| **CZ-2** | 0.010 | $-3.58 \pm .07$ | 5 |
| **CZ-3** | 0.001 | $-3.59 \pm .07$ | 5 |
| Stillinger-Weber Potential | | | |
| **Si-1** | 1.00 | $-4.09 \pm .16$ | 4 |
| **Si-2** | 0.10 | $-4.10 \pm .14$ | 5 |
| **Si-3** | 0.01 | $-4.11 \pm .13$ | 5 |

Table 3 – Mean and standard deviation for SBSR and STZ-PF model parameters and R-squared values. Average values are computed using fit data from N simulations. All lengths are made unitless by dividing by system height, L. Data reported for simulations where band width is bound and less than the system height.

|  |  | SBSR Model |  | STZ-PF Model |  |  |
| --- | --- | --- | --- | --- | --- | --- |
| Configuration | N | $\overline{\mathcal{L}}/L$ | $R^2$ | $\overline{\mathcal{L}}/L$ | $\overline{w_\infty}/L$ | $R^2$ |
| **LJ-1** | 6 | 0.018 ± 0.003 | 0.89 ± 0.08 | 0.034 ± 0.003 | 0.41 ± 0.03 | 0.99 ± 0.01 |
| **LJ-2** | 7 | 0.0069 ± 0.0008 | 0.94 ± 0.03 | 0.012 ± 0.001 | 0.27 ± 0.02 | 0.99 ± 0.01 |
| **LJ-3** | 7 | 0.0031 ± 0.0005 | 0.83 ± 0.19 | 0.006 ± 0.001 | 0.21 ± 0.16 | 0.97 ± 0.02 |
| **CZ-1** | 3 | 0.028 ± 0.002 | 0.98 ± 0.02 | 0.042 ± 0.005 | 0.69 ± 0.18 | 0.994 ± 0.005 |
| **CZ-2** | 4 | 0.0105 ± 0.0006 | 0.985 ± 0.004 | 0.014 ± 0.002 | 0.49 ± 0.15 | 0.995 ± 0.003 |
| **CZ-3** | 3 | 0.0019 ± 0.0003 | 0.70 ± 0.23 | 0.0037 ± 0.0004 | 0.138 ± 0.009 | 0.96 ± 0.03 |
| **Si-1** | 2 | 0.0108 ± 0.0007 | 0.9671 ± 0.0004 | 0.017 ± 0.002 | 0.56 ± 0.19 | 0.988 ± 0.006 |
| **Si-2** | 5 | 0.0060 ± 0.0003 | 0.984 ± 0.007 | 0.0083 ± 0.0006 | 0.41 ± 0.05 | 0.997 ± 0.002 |
| **Si-3** | 3 | 0.0033 ± 0.0002 | 0.97 ± 0.01 | 0.0048 ± 0.0003 | 0.31 ± 0.03 | 0.997 ± 0.001 |

# Figures

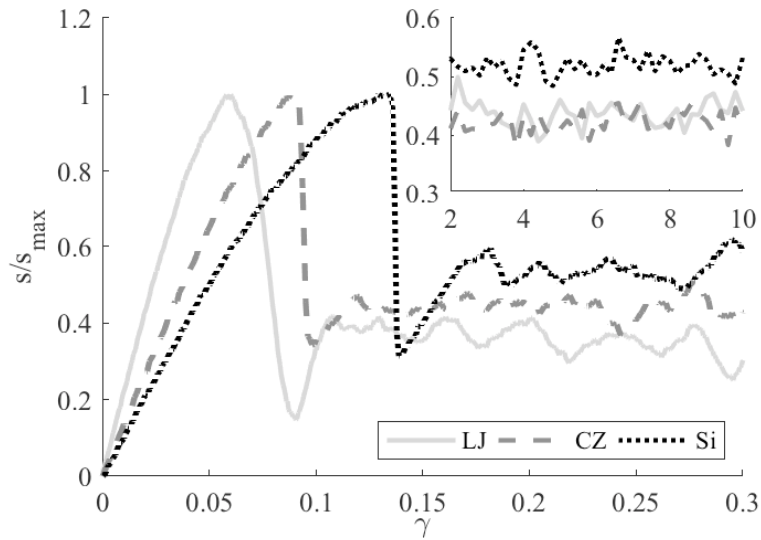

Figure 1 – Representative stress-strain curves for LJ (solid), CZ (dashed) and Si (dotted) systems. Shear stress has been normalized by the maximum stress and presented for strain from 0 – 30% (main) and 100% to 1000% (inlay). Data taken from a single simulation of the slowest-quenched configuration for a given system.

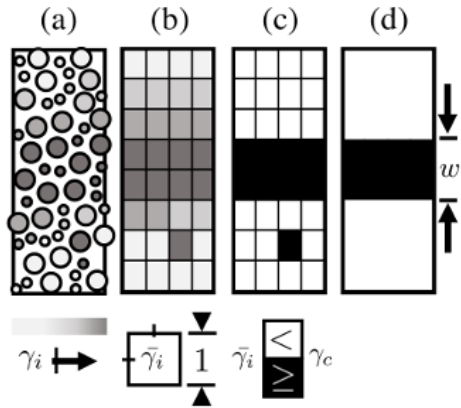

Figure 2 – Procedure for measuring shear band width. First we calculate per atom atomic strain $\gamma_i$, where $i$ indexes over all particles (a). Strain is then averaged in regions of square length to determine $\bar{\gamma}_i$ (b). A binary mask is applied with cutoff $\gamma_c = 0.25$ and $\bar{\gamma}_i < \gamma_c \to 0$ and $\bar{\gamma}_i \geq \gamma_c \to 1$ (c). Contiguous squares are grouped as features with special attention paid to features which lie on the edges of the simulation cell. We treat the largest feature as the shear band and measure its height in the y-direction to determine the bandwidth $w$ as a function of strain (d).

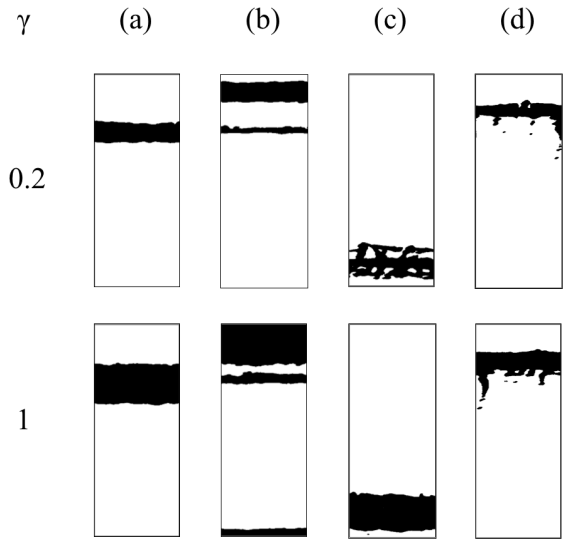

Figure 3 – High-strain material regions for multiple simulations when global strain is 20% (top row) and 100% (bottom row). Material regions with average atomic strain equal to or greater than 25% shown in black. (a) is representative of the ideal case where a single, horizontal shear band forms. In (b) two shear bands are present, disqualifying this simulation from our analysis. The dominant feature in (c) has a pore-like structure, with intermixed low and high-strain material regions. The high-strain region of (d) features a large vertical component.

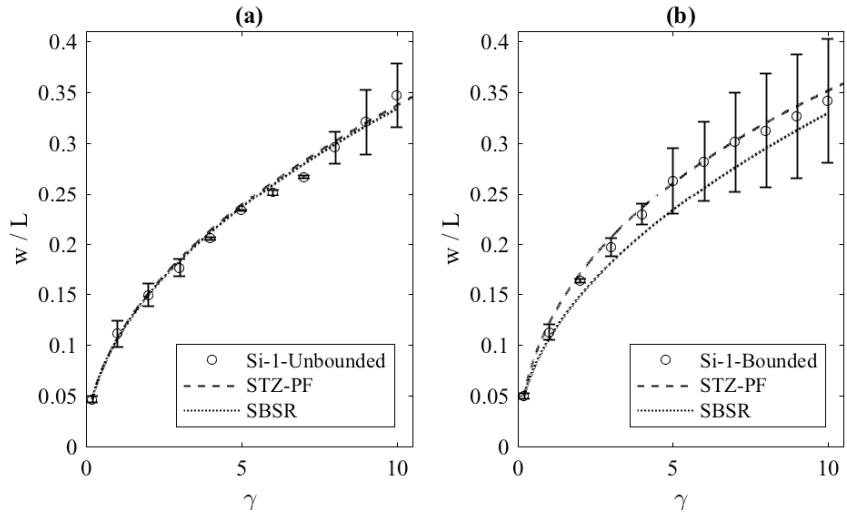

Figure 4 - Shear band width normalized by simulation height as a function of global strain for bounded (a) and unbounded (b) Si-1 configuration simulations. A shear band is considered bounded if the STZ-PF model fit yields a saturation width $w_\infty$ that is less than the height of the simulation cell in the direction perpendicular to the applied strain, $L$, and unbounded otherwise. Standard deviation bars are shown. Average model parameters are calculated from fits to individual simulations and used to generate the STZ-PF (dashed) and SBSR (dotted) curves.

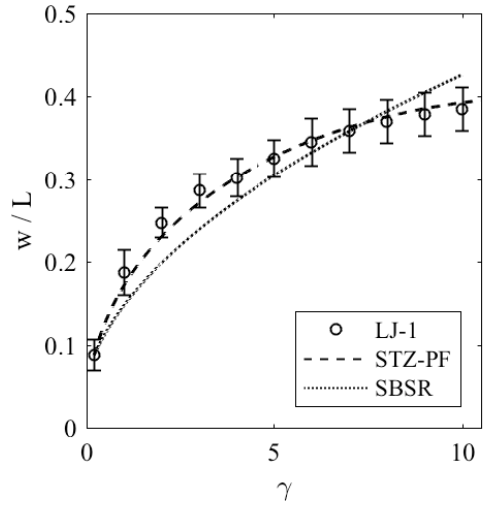

Figure 5 – Shear band width normalized by simulation height as a function of strain for LJ-1 configuration. Data averaged over 6 simulations where band growth is constrained and $w_\infty < L$ in the STZ-PF model. Standard deviation bars shown. Simulations are independently fitted to STZ-PF and SBSR models and mean fit parameters are computed. The resultant STZ-PF (dashed) and SBSR (dotted) fits are plotted.

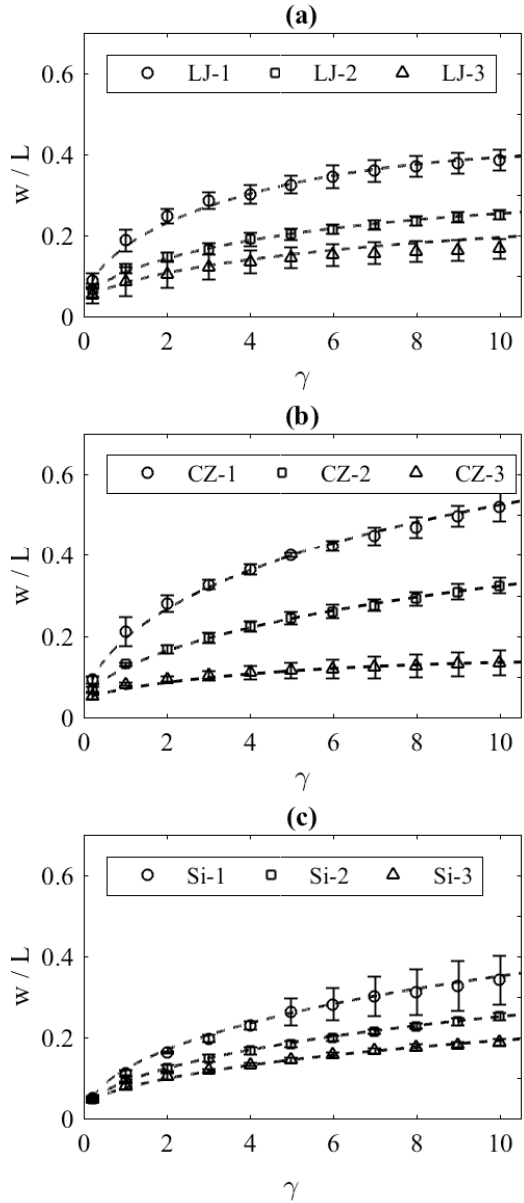

Figure 6 – Band width normalized by simulation height as a function of strain for LJ (a), CZ (b) and Si (c) systems. Data averaged over simulations where band growth is constrained and $w_\infty <  L$ in the STZ-PF model. Standard deviation bars shown. Circles represent fastest quench and triangles the slowest. STZ-PF model fit found by averaging over L and $w_\infty$ values and shown by dashed line.